\documentclass[apjl]{emulateapj}  
\usepackage{apjfonts}
\usepackage{mathptmx}

\shorttitle{Evershed clouds as precursors of moving magnetic features} 
\shortauthors{CABRERA SOLANA ET AL.}
\slugcomment{Received 2006 May 17; accepted 2006 August 7}
\journalinfo{The Astrophysical Journal, 649: L000-L000, 2006 September 20}
\begin{document}

\title{Evershed clouds as precursors of moving magnetic features around 
sunspots}
\author{D.\ Cabrera Solana\altaffilmark{1}, L.R.\ Bellot
 Rubio\altaffilmark{1}, C.\ Beck\altaffilmark{2}, and J.C.\ del Toro
 Iniesta\altaffilmark{1} }

\altaffiltext{1}{Instituto de Astrof\'{\i}sica de Andaluc\'{\i}a
(CSIC), Apdo.\ de Correos 3004, E-18080 Granada, Spain;
dcabrera@iaa.es; lbellot@iaa.es; jti@iaa.es} 
\altaffiltext{2}{Kiepenheuer-Institut f\"ur
Sonnenphysik, Sch\"oneckstr.\ 6, 79104, Freiburg, Germany; 
cbeck@kis.uni-freiburg.de}

\begin{abstract} 
The relation between the Evershed flow and moving magnetic features
(MMFs) is studied using high-cadence, simultaneous spectropolarimetric
measurements of a sunspot in visible (630.2~nm) and near-infrared
(1565~nm) lines. Doppler velocities, magnetograms, and
total linear polarization maps are calculated from the observed Stokes
profiles. We follow the temporal evolution of two Evershed clouds that
move radially outward along the same penumbral filament. Eventually,
the clouds cross the visible border of the spot and enter the moat
region, where they become MMFs. The flux patch farther from the
sunspot has the same polarity of the spot, while the MMF closer to it
has opposite polarity and exhibits abnormal circular polarization
profiles. Our results provide strong evidence that at least some MMFs
are the continuation of the penumbral Evershed flow into the moat.
This, in turn, suggests that MMFs are magnetically connected to
sunspots.
\end{abstract}


  \section{Introduction}
  \label{sec:intro}

Moving magnetic features (MMFs) are small flux concentrations observed
in the moat surrounding sunspots. They were discovered by Sheeley
(1969). Not surprisingly, a connection between MMFs and the magnetic
field of sunspots was proposed a few years later by Harvey \& Harvey
(1973). The phenomenological properties of MMFs are relatively well
know by now (Shine \& Title 2001), but their origin and magnetic
structure remain a mystery. The situation is more favorable in the
case of sunspots (Solanki 2003). According to current views, sunspot
penumbrae are formed by at least two magnetic components with
different field inclinations. The more horizontal component drives the
Evershed flow from the inner to the outer penumbra, where it partly
dives down below the photosphere or continues into the canopy. The
Evershed flow is not steady.  Shine et al.\ (1994), Rimmele (1994),
and Rouppe van der Voort (2003) have demonstrated the existence of
velocity packages, called ''Evershed clouds'', that propagate
radially outward and show an irregular repetitive behavior on a 
timescale of the order of 10--15~minutes. Evershed clouds evolve and remain
coherent until they go through the outer penumbral border, where they
seem to vanish.

Some investigators have pointed out the possible link between MMFs and
the penumbral component carrying the Evershed flow (Mart\'{\i}nez
Pillet 2002; Schlichenmaier 2002; Thomas et al.\ 2002). In fact,
recent observations seem to support this idea. Using Michelson Doppler
Imager magnetograms,
Sainz Dalda \& Mart\'{\i}nez Pillet (2005) discovered magnetic
filaments extending from the midpenumbra into the moat. Several
bipolar MMFs were observed to leave the sunspot following the paths
traced by these filaments. Zhang et al.\ (2003) and Penn et
al.\ (2006) also detected the passage of MMFs through the outer
penumbral boundary and their subsequent evolution across the
moat. Bonet et al.\ (2004) applied a phase diversity restoration to
G-band filtergrams of a sunspot. They concluded that most G-band
bright points, which are associated with MMFs (Beck et al.\ 2006), are
born close to the continuation of dark penumbral filaments. Hence,
they pointed out the possible relationship between G-band bright
points (or MMFs) and the more horizontal component of the
penumbra. Finally, Kubo (2005) analyzed Advanced Stokes Polarimeter
measurements. He found significant correspondence between the magnetic
field of MMFs and the penumbral uncombed structure.

In this Letter, the connection between the Evershed flow and MMFs is
established. Using high resolution spectropolarimetric measurements in
two different spectral ranges, we track the evolution of Evershed
clouds from the midpenumbra to well outside the outer penumbral
boundary, demonstrating that they become MMFs once they leave the
sunspot.

\section{Observations and data analysis}
\label{sec:obs}

\begin{figure}
\centering
\epsscale{1.11}
\plotone{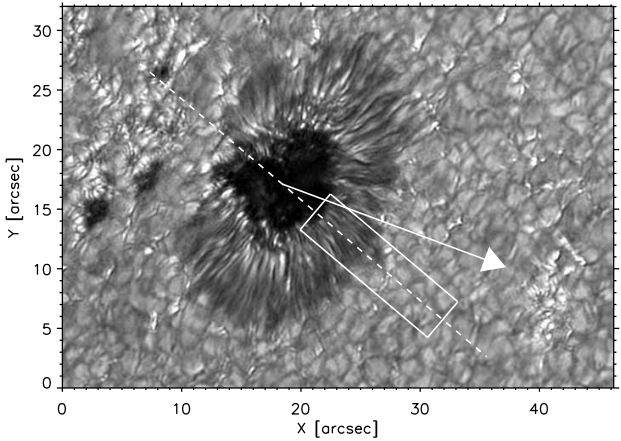}
\caption{Speckle-reconstructed G-band filtergram of AR 10781 taken at 
the Dutch Open Telescope (DOT) on 2005 June 30, 9:34 UT. The box shows the
fraction of the center-side penumbra and adjacent moat scanned by 
TIP and POLIS. The arrow marks the direction toward disk center.}
\label{fig:fig1}
\end{figure}

NOAA AR 10781 was observed on 2005 June 30 from 8:47 to
11:23 UT with the German Vacuum Tower Telescope (VTT) of Observatorio
del Teide (Tenerife, Spain) at an heliocentric angle of
$43^\circ$. The Tenerife Infrared Polarimeter (TIP; Mart\'{\i}nez
Pillet et al.\ 1999) and the POlarimetric LIttrow Spectrograph (POLIS;
Beck et al.\ 2005a) were operated simultaneously to record the full
Stokes profiles of the \ion{Fe}{1} lines at 1564.8, 1565.2,
630.15, and 630.25~nm. The slit width was 0\farcs36 for TIP and
0\farcs18 for POLIS. We performed rapid scans of a small portion of
the center-side penumbra of the spot, including the adjacent moat
region (see Figure~\ref{fig:fig1}). The scan step was 0$\farcs$2 for a
total of 20 steps (4$\arcsec$). The integration time was 10.5~s per
slit position, resulting in a cadence of 3.9 minutes. We performed 40
repetitions of the scan. During the observations, the Kiepenheuer
Adaptive Optics System (Soltau et al.\ 2002) was used to reduce image
motion and blurring. This, together with excellent seeing conditions,
allowed us to reach a spatial resolution of about
0\farcs6--0\farcs7. The data sets have been coaligned by
cross-correlation using the 0$\farcs$175 $\times$ 0$\farcs$175 TIP
pixel as a reference (Beck et al.\ 2006).

After correcting the measurements for instrumental polarization (Beck
et al.\ 2005b), residual crosstalk is estimated to be on the order of
$10^{-3}$ in units of the continuum intensity. The velocity scale for
TIP has been set calculating a mean quiet Sun intensity profile. The
wavelengths of the line cores in the mean profile, corrected for
convective blueshifts (Borrero \& Bellot Rubio 2002), represent the
position of zero velocity. For POLIS, we have used the teluric O$_2$
lines to perform an absolute velocity calibration for each repetition
of the scan in the same way as Mart\'inez Pillet et al.\ (1997).

\begin{figure*}
\centering
\epsscale{1.1}
\plotone{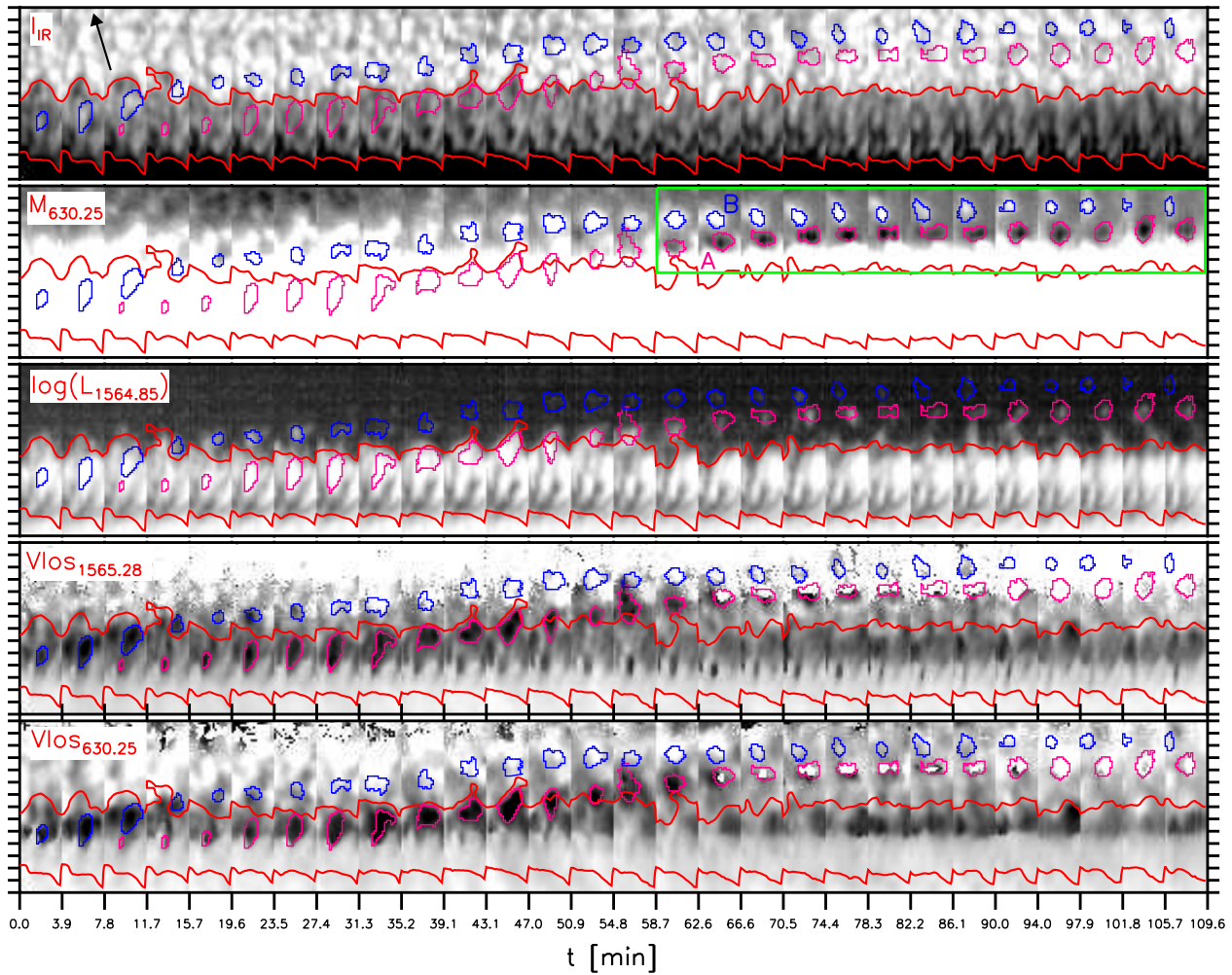}
\caption{From {\em top} to {\em bottom}: Continuum intensity at 1565
nm, magnetogram signal from \ion{Fe}{1} 630.25~nm, logarithm of the
total linear polarization of \ion{Fe}{1} 1564.85~nm, Stokes $V$
zero-crossing velocity of \ion{Fe}{1} 1565.28~nm, and Stokes $V$
zero-crossing velocity of \ion{Fe}{1} 630.25~nm. The gray scale ranges
from 0.82 to 0.95 in the first panel, from $-0.5\%$ to 1\% in the
second panel, from $-0.97$ to 0.80 in the third panel, from $-4.0$ to
0~km~s$^{-1}$ in the fourth panel, and from $-2.8$ to 0~km~s$^{-1}$ in
the last panel. Negative velocities are blueshifts. Pink and blue
contours delimit features A and B, respectively. Red lines indicate
the inner and outer penumbral boundaries. Each tickmark in the
$y$-axis represents 1\arcsec.  The arrow marks the direction toward disk
center. $t=0$ minutes corresponds to June 30, 9:34 UT.}
\label{fig:fig2}
\end{figure*}

Line-of-sight (LOS) velocities and magnetograms have been calculated
from the observed circular polarization profiles $V(\lambda)$. LOS
velocities are estimated from the Doppler shifts of the Stokes $V$
zero-crossing wavelengths. These velocities are representative of the
magnetic atmosphere inside the resolution element. Additionally, we
construct magnetograms, $M$, as (minus) the Stokes $V$ signal of
\ion{Fe}{1} 630.25 nm at $\Delta \lambda = +10$~pm from line
center. $M$ provides a rough estimate of the longitudinal magnetic
flux.  The total linear polarization, $L = \sum_{i}
[Q(\lambda_{i})^2 + U(\lambda_{i})^2]^{1/2}$, is also calculated. For
weak lines, $L$ is proportional to $\sin^2{\gamma}$ if the magnetic
field orientation remains constant along the LOS. Hence, the larger
the value of $L$, the greater the field inclination to the LOS.

\section{Results}
\label{sec:res}

In Figure~\ref{fig:fig2} we show the temporal evolution of the
center-side penumbra and moat region studied in this work.  The first
panel displays the continuum intensity at 1565 nm. The next two panels
show the magnetograms derived from \ion{Fe}{1} 630.25~nm and the total
linear polarization of the \ion{Fe}{1} line at 1564.85~nm. The last
two panels display the LOS velocities calculated from the Stokes $V$
zero-crossing shifts of \ion{Fe}{1} 1565.28 and \ion{Fe}{1}
630.25~nm. The velocity is computed only for pixels exhibiting normal
two-lobed $V$ profiles with amplitudes greater than 0.2$\%$ for
\ion{Fe}{1} 630.25~nm and 0.5$\%$ for \ion{Fe}{1} 1564.85~nm.

During the time sequence, two opposite-polarity flux patches are
observed to move away from the spot in the moat region (cf.\ the area
enclosed by the green box in the magnetograms of
Figure~\ref{fig:fig2}). We have outlined them with pink and blue
contours, and will refer to them as features A and B,
respectively. The properties of this pair are similar to those of
bipolar MMFs: the separation between the two polarities is around
2$\arcsec$ (Zhang et al.\ 2003), the magnetic patch having the same
polarity as the sunspot is located farther from the spot (Lee 1992;
Yurchyshyn et al.\ 2001; Zhang et al.\ 2003), and the average
proper motion, corrected for the viewing angle, is around
0.5~km~s$^{-1}$ (Nye et al.\ 1984; Zhang et al.\
2003). Hence, features A and B could be classified as a bipolar MMF
or, according to Shine \& Title (2001), as a type I MMF. However, from
the magnetogram data alone we cannot rule out the possibility that A
and B are two unrelated MMFs of type II and III. In any case, the
second and third panels of Figure~\ref{fig:fig2} demonstrate that the
two flux patches possess different field inclinations.  While feature
A shows up in both $L$ and $M$, feature B is observed mainly in
$M$. This indicates that A has more inclined magnetic fields to the
LOS than B.

Our high-cadence observations make it possible to investigate the
formation of these MMFs. Going backwards in time, one can see in
the velocity maps that the origin of feature A is an Evershed cloud
that shows up in the midpenumbra at $t=7.8$ minutes.  The cloud is
detected as a region of increased velocity signals, with maximum
blueshifts of 3.2~km~s$^{-1}$ in \ion{Fe}{1} 630.25~nm and
4.1~km~s$^{-1}$ in \ion{Fe}{1} 1565.28~nm. Initially its spatial extent
is 750 km, but as time goes on it becomes longer toward the outer
penumbral boundary, reaching 2250 km at $t=31.3$ minutes. Near the sunspot
edge, the cloud adopts a roundish shape ($t=35.2$--43.1 minutes).  The
third panel of Figure~\ref{fig:fig2} demonstrates that this structure
moves radially outward following a penumbral filament with larger
values of the linear polarization (that is, larger inclinations of the
magnetic field to the LOS) than its surroundings. Its average
propagation speed inside the penumbra is $\sim 1.7$~km~s$^{-1}$. The
Evershed cloud crosses the penumbral boundary between $t= 47.0$ and
$t= 50.9$ minutes. Outside the spot, it is observed both in the velocity
maps and in the magnetograms as a negative polarity MMF.

The precursor of feature B is another Evershed cloud first detected at
$t=0$ minutes in the midpenumbra. It propagates along the very same
penumbral filament as feature A, crossing the outer boundary of the
spot at $t= 11.7$ minutes. The cloud moves with an average speed of
3.3~km~s$^{-1}$ inside the penumbra and exhibits maximum blueshifts of
3.2~km~s$^{-1}$ (\ion{Fe}{1} 630.25~nm) and 4.4~km~s$^{-1}$ (\ion{Fe}{1}
1565.28~nm). Once in the moat region, it is still observed as a
velocity structure until $t= 27.4$ minutes, and then it shows up in the
magnetograms as a positive polarity MMF.

\section{Discussion}
As it would be the case with any magnetograph observation of features
A and B (regardless of the instrument used or the spatial resolution),
the magnetograms of Figure~\ref{fig:fig2} provide little
information on their magnetic structure for two reasons.

First, the observed polarities do not reflect true {\em field 
inclinations} because of projection effects (the sunspot was located 
$43^{\circ}$ off the disk center). To derive true inclinations one 
has to infer the vector magnetic field in the LOS frame and transform 
it to the local reference frame, which can only be done through 
inversions of all four Stokes spectra. 

\begin{figure*}
\centering
\epsscale{1.15}
\plotone{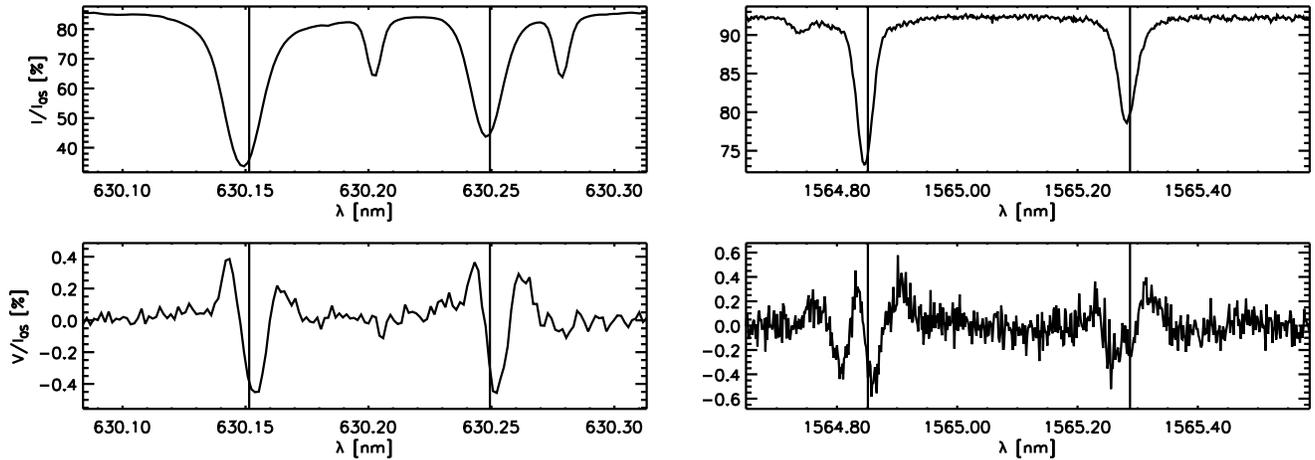}     
\caption{Cospatial Stokes $I$ ({\em top}) and $V$ ({\em bottom})
profiles of \ion{Fe}{1} 630.15, \ion{Fe}{1} 630.25,
\ion{Fe}{1} 1564.85, and \ion{Fe}{1} 1565.2 emerging from a
pixel inside feature A with abnormal circular polarization
signals.  The vertical solid lines represent the rest
wavelengths of the spectral lines.}
\label{fig:fig4}
\end{figure*}

Second, and more important, abnormal Stokes $V$ profiles with three or
four lobes occur frequently in MMFs. This is demonstrated for the
first time by our observations. In Figure~\ref{fig:fig4} we show an
example of cospatial Stokes $I$ and $V$ profiles of \ion{Fe}{1}
630.15, \ion{Fe}{1} 630.25, \ion{Fe}{1} 1564.85, and
\ion{Fe}{1} 1565.2~nm emerging from a negative-polarity pixel inside
feature A. The very complex shapes displayed by the Stokes $V$
profiles of the four lines may be caused by strong gradients of the
atmospheric parameters along the LOS and/or the presence of several
magnetic components in the resolution element. The problem with such
abnormal profiles is that the meaning of polarity looses its sense, as
both positive and negative polarities can be assigned to the same
pixel depending on the exact wavelength(s) used to construct the
magnetogram.  It is not surprising that the existence of abnormal
profiles has gone unnoticed: most of what we know today about MMFs has
been learned from magnetograph observations that cannot reveal this
kind of spectral subtleties. One should realize, however, that gross
errors may result from direct interpretations of magnetograms when
abnormal profiles exist, simply because even the polarity of the field
would be an ill-defined quantity. This is the reason why we refrain
from attempting a definite classification of the observed MMFs. The
conclusion that MMFs can be traced back to Evershed clouds appearing
in the midpenumbra, however, is a solid result that does not depend
in any way on the exact nature of the flux concentrations.
%

\section{Conclusions}
\label{sec:con}

We have used simultaneous spectropolarimetric measurements in two
spectral regions to demonstrate the relation between the penumbral
Evershed flow and moving magnetic features (MMFs). Our observations
combine high cadence (3.9 minutes) with one of the highest spatial 
resolutions achieved in solar spectropolarimetry ($\sim 0\farcs7$).

One after the other, two Evershed clouds are seen to move radially
outward along a penumbral filament characterized by larger field
inclinations to the LOS than its surroundings. They appear in the
center-side penumbra, exhibiting blueshifts and positive polarity
(i.e., the polarity of the spot). Eventually, the clouds cross the
outer penumbral boundary and become MMFs. Once in the moat region, the
flux patch closer to the spot shows negative polarity. The two MMFs
are separated by about 2\arcsec and move away from the spot with an
average speed of 0.5~km~s$^{-1}$. We have examples of other Evershed
clouds crossing the outer penumbral boundary and becoming MMFs, so the
process just described may be quite common in sunspots.

Our observations demonstrate that at least some MMFs are a
manifestation of the penumbral Evershed flow in the sunspot
moat. This, in turn, gives support to earlier results about the
magnetic connection between MMFs and the penumbra (e.g., Zhang et al.\
2003; Sainz Dalda \& Mart\'{\i}nez Pillet 2005). The association
between Evershed clouds and MMFs found here suggests that the physical
structures responsible for the existence of the former inside the
penumbra are also present in, or propagate to, the sunspot moat. This
may help refine theoretical models of sunspot penumbrae.

We have also shown, for the first time, that abnormal circular
polarization profiles with three or four lobes occur in MMFs. The
presence of abnormal profiles hampers direct interpretations of
magnetogram polarities in terms of magnetic field inclinations. As a
result, we do not know if the vector magnetic field is really pointing
downward in the negative-polarity MMF and upward in the the
positive-polarity MMF, or if the field azimuth provides any evidence
of magnetic connection between the two flux concentrations. To
investigate these issues, our efforts will concentrate on the
inversion of the observed Stokes spectra.

\acknowledgments Discussions with J.M.\ Borrero and A.\ Sainz Dalda
are gratefully acknowledged. The DOT G-band image was taken from
http://dotdb.phys.uu.nl/DOT/.  This work has been supported by the
Spanish MEC through {\em Programa Ram\'on y Cajal} and projects
ESP2002-04256-C04-01 and ESP2003-07735-C04-03, and by the German
DFG through grant SCHL 514.  The VTT is operated by
the Kiepenheuer-Institut f\"ur Sonnenphysik at the Spanish 
Observatorio del Teide of the Instituto de Astrof\'{\i}sica de 
Canarias.

\end{document}